## SPECTROSCOPY AND ORBITAL ANALYSIS OF BRIGHT BOLIDES OBSERVED OVER THE IBERIAN PENINSULA FROM 2010 TO 2012

# SPECTROSCOPY AND ORBITAL ANALYSIS OF BRIGHT BOLIDES OBSERVED OVER THE IBERIAN PENINSULA FROM 2010 TO 2012

José M. Madiedo[1,2], Josep M. Trigo-Rodríguez[3], José L. Ortiz[4], Alberto J. Castro-Tirado[4], Sensi Pastor[5], José A. de los Reyes[5] and Jesús Cabrera-Caño[2]

[1] Facultad de Ciencias Experimentales, Universidad de Huelva, 21071 Huelva, Spain.

[2] Departamento de Física Atómica, Molecular y Nuclear. Facultad de Física. Universidad de Sevilla. 41012 Sevilla, Spain.

[3] Institut de Ciències de l'Espai (CSIC-IEEC), Campus UAB, Facultat de Ciències, Torre C5-parell-2ª, 08193 Bellaterra, Barcelona, Spain.

[4] Instituto de Astrofísica de Andalucía, CSIC, Apt. 3004, 18080 Granada, Spain.

[5]Observatorio Astronómico de La Murta. Molina de Segura, 30500 Murcia, Spain.

## ABSTRACT

We present the analysis of the atmospheric trajectory and orbital data of four bright bolides observed over Spain, one of which is a potential meteorite dropping event. Their absolute magnitude ranges from -10 to -11. Two of these are of sporadic origin, although a Geminid and a κ-Cygnid fireball are also considered. These events were recorded in the framework of the continuous fireball monitoring and spectroscopy campaigns developed by the SPanish Meteor Network (SPMN) between 2010 and 2012. The tensile strength of the parent meteoroids is estimated and the abundances of the main rock-forming elements in these particles are calculated from the emission spectrum obtained for three of these events. This analysis revealed a chondritic nature for these meteoroids.

## 1 INTRODUCTION

The detailed study of bright fireballs is one of the aims of the SPanish Meteor Network (SPMN), as these allow us to collect very valuable information on the origin and properties of large meteoroids that, under appropriate conditions, can give rise to the relatively rare meteorite-dropping events. Thus, one of the main goals of our meteor network is the analysis of the physico-chemical properties of these meteoroids from multiple station images. For this purpose we perform a systematic monitoring of the night sky with the aim to obtain data to improve our knowledge of the mechanisms that deliver these materials to the Earth. Nowadays we operate 25 meteor observing stations that monitor the night sky over Spain and neighbouring regions, which is equivalent to an area of more than 500.000 km$^2$. These stations provide useful information for the determination of radiant, orbital and photometric parameters (Madiedo and Trigo-Rodriguez 2008; Trigo-Rodriguez et al. 2007, 2009a; Madiedo et al. 2013a, 2013b, 2013c). Besides, we focus special attention on the study of the chemical composition of meteoroids from the analysis of the emission spectrum produced when these particles ablate in the atmosphere (Trigo-Rodriguez et al 2007, 2009a; Madiedo et al. 2013a, 2013b, 2013c). For this reason, a significant part of our effort has focused on the deployment of spectrographs at some of our meteor observing stations. The first of these devices, which were based on high-sensibility CCD video cameras endowed with attached holographic diffraction gratings, started operation in 2006 from our station in Sevilla, but also from a mobile station operated when necessary from Cerro Negro, a dark countryside environment at about 60 km north from Sevilla. Thus, we are performing from Sevilla a continuous spectroscopic campaign since that year. Nowadays, these spectral video cameras work continuously from 8 SPMN stations. Favourable weather conditions in Spain play a key role in the successful development of this spectroscopic campaign. In this context, we present here the analysis of four bolides with a minimum brightness equivalent to an absolute magnitude of -10. These were recorded over the Iberian Peninsula in the framework of our continuous fireball monitoring and meteor spectroscopy campaigns performed between 2010 and 2012.

## 2 INSTRUMENTATION AND DATA REDUCTION TECHNIQUES

High-sensitivity monochrome CCD video cameras (models 902H2 and 902H Ultimate from Watec Corporation, Japan) were employed to image the fireballs discussed here. Thus, the bolides were recorded by an array of these video devices operating at the meteor observing stations listed in Table 1. Some of these stations work in an autonomous way by means of software developed by us (Madiedo & Trigo-Rodríguez 2010; Madiedo et al. 2010) and the cameras are arranged in such a way that the common atmospheric volume monitored by neighbouring stations is maximized. These devices generate interlaced imagery according to the PAL video standard. Thus, they generate video files at a rate of 25 frames per second and with a resolution of 720x576 pixels. Aspherical fast lenses with focal lengths ranging from 6 to 20 mm and focal ratios between 1.4 and 0.8 were used for the imaging objective lens. In this way, different areas of the sky were covered by every camera and point-like star images were obtained across the entire field of view. With this configuration we can image meteors with an apparent magnitude of about 3±1. A more detailed description of the operation of these systems is given in (Madiedo & Trigo-Rodríguez 2008, 2010).

For meteor spectroscopy we employ holographic diffraction gratings (500 or 1000 lines/mm, depending on the device) attached to the objective lens of some of the above-mentioned cameras to image the emission spectra resulting from the ablation of meteoroids in the atmosphere. We do not employ additional image intensifying devices. We can image spectra for meteor events with brightness higher than mag. -3/-4.

With respect to data reduction, once that meteor trails simultaneously recorded from at least two different meteor stations are identified, we first deinterlaced the video images provided by our cameras. Thus, even and odd fields were separated for each video frame, and a new video file containing these was generated. Since this operation implies duplicating the total number of frames, the frame rate in the resulting video is 50 fps. Then, an astrometric measurement is done by hand in order to obtain the plate (x, y) coordinates of the meteor along its apparent path from each station. The astrometric measurements are then introduced in our AMALTHEA software (Trigo-Rodríguez et al. 2009a; Madiedo et al. 2011a), which transforms plate coordinates into equatorial coordinates by using the position of reference stars appearing in the images. This package employs the method of the intersection of planes to determine the position of the apparent radiant and also to reconstruct the trajectory in the atmosphere of meteors recorded from at least two different observing stations (Ceplecha 1987). In this way, the beginning and terminal heights of the meteor are inferred. From the sequential measurements of the video frames and the trajectory length, the velocity of the meteor along its path is obtained. The preatmospheric velocity $V_\infty$ is found from the velocity measured in the earliest parts of the meteor trajectory. Once these data are known, the software computes the orbital parameters of the corresponding meteoroid by following the procedure described in Ceplecha (1987).

### 3 OBSERVATIONS: atmospheric trajectory, radiant and orbit

The fireballs analyzed here are listed in Table 2. Their SPMN code, which was assigned after the recording date, is included for identification. The times of fireball passage are known with a precision of 0.1 seconds. Besides, the bolides were named according to the geographical location nearest to the projection on the ground of their atmospheric trajectory. The beginning ($H_b$), ending ($H_e$) and maximum brightness ($H_{max}$) heights are also indicated. As can be noticed, their absolute magnitude (M) ranges from -10.0±0.5 to -11.0±0.5. These magnitudes were calculated by direct comparison of the brightness level of the pixels near the maximum luminosity of the meteor trail and those of nearby stars. Each bolide was simultaneously recorded from, at least, two of the observing stations listed in Table 1. The radiant and orbital data obtained from the analysis of these events are summarized in Table 3. These bolides are described in more detail below.

### 3.1 The "La Carlota" fireball (SPMN180910)

This sporadic fireball, which reached the end of its luminous phase next to the zenith of the city after which it was named, was simultaneously imaged from stations #1 and #9 on September 18, 2010 (Figure 1). The event, with an estimated absolute magnitude of -11.0±0.5, began at 20h04m27.0±0.1s UTC at a height of 85.8±0.5 km above the ground level and lasted about 4.5 seconds. The preatmospheric velocity of the meteoroid was 21.7±0.3 km s$^{-1}$, with the apparent radiant located at $\alpha$=300.8±0.7 °, $\delta$=45.9±0.3 °. The bolide penetrated the atmosphere till a height of 25.6±0.5 km. The projection on the ecliptic plane of the heliocentric orbit of the meteoroid is shown in Figure 1d. The calculated orbital period yields P=5.9±0.5 yr, and the value obtained for the Tisserand parameter with respect to Jupiter is $T_J$=2.5±0.1. This reveals that the particle was following a Jupiter Family Comet (JFC) orbit (orbital period P < 20 yr and Tisserand parameter in the range 2 <$T_J$ < 3) before impacting the Earth.

### 3.2 The "Doñana" fireball (SPMN250112)

A mag. -10.1±0.5 slow-moving fireball was simultaneously recorded from stations #1 and #5 (Sevilla and El Arenosillo) on January 25, 2012, at 20h20m07.3±0.1s UTC (Figure 2). The apparent trajectory of this sporadic bolide, which lasted about 6 seconds, is shown in Figure 4c. The parent meteoroid struck the atmosphere with an initial velocity $V_\infty$=14.7±0.3 km s$^{-1}$ and a zenith angle of 38.2 º, with an apparent radiant located at α=42.3±0.3 º, δ=3.8±0.3 º. The luminous phase began at 78.4±0.5 km above the ground level and ended at 26.4±0.5 km. This deep-penetrating fireball received the name "Doñana", as its atmospheric path was located over this natural park in the south of Spain. Once this trajectory was obtained, the orbital parameters of the meteoroid were calculated (Table 3). The projection of this orbit on the ecliptic plane is plotted in Figure 2d. The Tisserand parameter with respect to Jupiter ($T_J$=2.4±0.1) and the orbital period (P=8.3±1.8 yr) show that this sporadic event was also produced by a meteoroid in a JFC orbit.

### 3.3 The "Torrecera" fireball (SPMN150812)

This bolide, which lasted about 2 seconds, was recorded from stations #1 and #5 on August 15, 2012, at 23h44m39.2±0.1s UTC (Figure 3). According to the photometric analysis of the images, the event reached an absolute magnitude of about -10.2±0.5. As can be seen in Figures 4a and 4b, the fireball experienced a very bright fulguration at the end of its luminous path as a consequence of the violent disruption of the parent meteoroid. According to our analysis, the fireball began at a height of 104.5±0.5 km and ended at 80.0±0.5 km above the ground level. The meteoroid struck the atmosphere with an initial velocity $V_\infty$=27.1±0.3 km s$^{-1}$ and the apparent radiant was located at α=295.4±0.3 º, δ=60.2±0.2 º. The projection on the ecliptic plane of the orbit of the meteoroid in the Solar System is shown in Figure 3d. These results confirm the association of this event with the kappa-Cygnid meteoroid stream.

### 3.4 The "Peñaflor" bolide (SPMN121212)

This mag. -10.0±0.5 Geminid fireball was simultaneously imaged from Sevilla, La Hita, and El Arenosillo on December 12, 2012 at 3h47m19.7±0.1s UTC (Figure 4). It was, in fact, the brightest Geminid recorded by our team in 2012. The analysis of its atmospheric path shows that the meteoroid struck the atmosphere with an initial velocity $V_\infty$=39.0±0.3 km s$^{-1}$ and a zenith angle of 13.4 º. The fireball, with an apparent radiant located at $\alpha$=116.2±0.3 º, $\delta$=35.1±0.3 º, began at 101.6±0.5 km above the ground level and ended at a height of 39.9±0.5 km. It lasted about 2.7 seconds. The heliocentric orbit of the meteoroid is shown in Figure 4d.

## 4 RESULTS AND DISCUSSION

### 4.1 Light curves and initial masses

The light curves (pixel intensity in arbitrary units vs. time) for the fireballs listed in Table 2 are shown in Figures 5 to 8. As can be noticed, all these events exhibit their maximum brightness during the second half of their trajectory, a behaviour which is typical of fireballs produced by compact meteoroids (Murray et al. 1999; Campbell et al. 2000). For dustball meteoroids, however, this takes place earlier (Hawkes & Jones 1975). In particular, the SPMN150812 "Torrecera" $\kappa$-Cygnid bolide (Figure 3) showed a very bright fulguration at the end of its atmospheric path as a consequence of the catastrophic disruption of the meteoroid.

The initial photometric mass $m_p$ of the parent meteoroids have been calculated from these light curves as the total mass lost due to the ablation process between the beginning of the luminous phase and the terminal point of the atmospheric trajectory of the bolides:

$$m_p = 2 \int_{t_b}^{t_e} I_p / (\tau v^2) dt \tag{1}$$

where $t_b$ and $t_e$ are, respectively, the times corresponding to the beginning and the end of the luminous phase. $I_p$ is the measured luminosity of the fireball, which is related to the absolute magnitude M by means of

$$M = -2.5 \log (I_p) \tag{2}$$

For the estimation of the luminous efficiency $\tau$, which depends on velocity, we have employed the equations given by Ceplecha & McCrosky (1976). The calculated values of the initial photometric mass for each bolide are listed in Table 4. Both "La Carlota" and "Doñana" sporadic bolides were produced by dm-sized meteoroids with masses of about 1.9 and 5.7 kg, respectively. The parent meteoroids of the "Fuencaliente" κ-Cygnid and the "Peñaflor" Geminid fireballs were, however, smaller. Their mass was around one tenth of the mass of the "Doñana" sporadic meteoroid (about 0.18 and 0.12 kg, respectively), despite their maximum luminosity was similar.

## 4.2 Tensile strength

As can be noticed in the composite images shown in Figures 1 to 4 and also in the above discussed light curves plotted in Figures 5 to 8, the fireballs exhibited at least one very bright flare along their atmospheric path. These events are typically produced by the fragmentation of the meteoroids when these particles penetrate denser atmospheric regions. Thus, once the overloading pressure becomes larger than the particle strength, the particle breaks apart. Quickly after that, a bright flare is produced as a consequence of the fast ablation of tiny fragments delivered to the thermal wave in the fireball's bow shock. The so-called tensile (aerodynamic) strength S at which these breakups take place can be calculated by means of the following equation (Bronshten 1981):

$$S = \rho_{atm} \cdot v^2 \qquad\qquad (3)$$

where v is the velocity of the meteoroid at the disruption point and $\rho_{atm}$ the atmospheric density at the height where the fracture takes place. The density can be calculated, for instance, by employing the US standard atmosphere model (U.S. Standard Atmosphere 1976). This aerodynamic strength S can be used as an estimation of the tensile strength of the meteoroid (Trigo-Rodriguez & Llorca 2006, 2007). The values obtained for the fireballs analyzed in this work are summarized in Table 5, together with the corresponding heights and velocities.

As Figure 5 shows, the SPMN180910 "La Carlota" fireball experienced its main fulguration at the instant $t_1 = 2.5$ s after the event started its luminous phase, followed by a second fulguration at the instant $t_2 = 3.3$ s. The images reveal that the corresponding fragmentations were not catastrophic, as the remaining material continued penetrating in the atmosphere. These took place, respectively, at a height of 51.3±0.5 and 38.1±0.5 km above the ground level. In this way, we infer that the meteoroid exhibited the first flare under a dynamic pressure of $(3.2\pm0.4)\cdot10^6$ dyn cm$^{-2}$, while the second flare considered here took place at $(1.4\pm0.4)\cdot10^7$ dyn cm$^{-2}$. The latter value is similar to the tensile strength found for stony meteorites (Consolmagno and Britt 1998; Consolmagno et al. 2006, 2008; Macke et al. 2011). This supports the idea of high-strength meteoroids moving in JFC orbits. Thus, although in general cometary meteoroids have low tensile strengths ranging from $\sim10^3$ to $\sim10^5$ dyn cm$^{-2}$ (Trigo-Rodríguez & Llorca 2006, 2007), fireballs produced by high-strength meteoroids moving in cometary orbits have also been reported. Examples of fireballs produced by high-strength meteoroids moving in cometary orbits are the Karlštejn fireball (Spurný & Borovička 1999a,b), for which a mechanical strength of about $7\cdot10^6$ dyn cm$^{-2}$ was obtained, and the deep-penetrating Bejar bolide (Trigo-Rodríguez et al. 2009b), for which the calculated tensile strength was $14\cdot10^7$ dyn cm$^{-2}$. Besides, there are evidences that even Leonid meteoroids, whose parent body is Comet 55P/Tempel–Tuttle, contain also much stronger ingredients with measured tensile strengths of about $2\cdot10^7$ dyn cm$^{-2}$ (Spurný et al. 2000, Borovička & Jenniskens 2000). Thus, some authors have proposed that cometary nuclei may also contain compact materials analogous to (or identical to) CI and CM carbonaceous chondrites (Lodders & Osborne 1999).

A similar behaviour is found for the SPMN250112 "Doñana" sporadic fireball (Figure 6), but also for the SPMN121212 "Peñaflor" Geminid bolide (Figure 8), in the sense that two main flares taking place at different heights can be distinguished. As can be seen in Table 4, for the "Doñana" event these flares took place under a dynamic pressure of $(3.6\pm0.4)\cdot10^6$ and $(5.6\pm0.4)\cdot10^6$ dyn cm$^{-2}$, respectively. So, this fireball was also produced by a tough meteoroid following a JFC orbit. On the other hand, the aerodynamic pressure obtained for the flares exhibited by the "Peñaflor" bolide $((2.0\pm0.4)\cdot10^5$ and $(5.1\pm0.4)\cdot10^5$ dyn cm$^{-2}$) are close to the tensile strength values previously obtained for other members of the Geminid meteoroid stream by Trigo-Rodríguez & Llorca (2006, 2007). The SPMN150812 "Torrecera" κ-Cygnid bolide, however, exhibited only one flare corresponding to the catastrophic disruption of the meteoroid at the end of its luminous phase (Figure 7). This break-up took place at a height of $80.0\pm0.5$ km. The calculation of the aerodynamic strength yields $(1.1\pm0.4)\cdot10^5$ dyn cm$^{-2}$. This value is of the same order of magnitude than those previously determined for the other κ-Cygnids (Trigo-Rodriguez et al. 2009a).

### 4.3 Emission spectra

Our spectral cameras obtained the emission spectrum of the above discussed fireballs, except for the SPMN180910 ("La Carlota") event. These can be used to obtain an insight into the chemical nature of the parent meteoroids. We have employed our CHIMET software to process these spectra (Madiedo et al. 2011b). This application follows the analysis procedure described in Trigo-Rodríguez et al. (2003). Thus, the video frames containing the emission spectrum were dark-frame substracted and flat-fielded. Then, the signal was calibrated in wavelengths by identifying typical lines appearing in meteor spectra (Ca, Fe, Mg and Na multiplets) and corrected by taking into account the efficiency of the recording instrument. The result can be seen in Figures 9 to 11. Multiplet numbers are given according to Moore (1945). For the "Peñaflor" and "Torrecera" bolides the emission spectrum is dominated by the lines of Mg I-2 (517.2 nm) and several Fe I multiplets: Fe I-41 (441.5 nm) and Fe I-15 (526.9 and 542.9 nm). In the ultraviolet, the intensity of H and K lines of ionized calcium are also strong, also blended with Fe I-4 due to the moderately low spectral resolution. The contribution from Na I-1 is also noticeable. The emission from atmospheric $N_2$ bands can also be noticed in the red region. The "Peñaflor" fireball, due to its higher velocity, exhibits a distinctive ionized Si II-2 line at 634.7 nm. On the other hand, the "Doñana" emission spectrum is characteristic of a low-velocity fireball mostly dominated by Fe I and Na I-1 lines and with a weaker Mg I-2 line than the "Torrecera" and "Peñaflor" events (Trigo-Rodríguez, 2002, Trigo-Rodríguez et al., 2003; 2004).

Once the main emission lines were identified, the relative abundances of the main rock-forming elements in the meteoroids were calculated. Thus, a software application was used to reconstruct a synthetic spectrum by adjusting the temperature (T) in the meteor plasma, the column density of atoms (N), the damping constant (D) and the surface area (P) from the observed brightness of lines as explained in Trigo-Rodríguez et al. (2003). First, as Fe I emission lines are well distributed all over the spectrum, they were used to set the parameters capable to fit the observed spectrum with the synthetic one (Trigo-Rodríguez et al. 2003, 2004). The modelled spectrum of the meteor column is then compared line by line with the observed one to fit all its peculiarities. Thus, once the parameters T, D, P and N are fixed, the relative abundances of the main rocky elements relative to Fe are found by performing an iterative process until an optimal fit is achieved. To obtain the elemental abundances relative to Si, we have considered Fe/Si=1.16 (Anders and Grevesse 1989). This was not needed for the Peñaflor bolide due to the possibility to obtain directly the Si abundance in the meteor column. The results are summarized in Table 6, where these relative abundances are compared with those found for other undifferentiated bodies in the Solar System, such as comet 1P/Halley and the CI and CM groups of carbonaceous chondrites (Jessberger et al. 1988; Rietmeijer & Nuth 2000; Rietmeijer 2002; Trigo-Rodríguez et al. 2003). Overall, the computed abundances suggest that the main rock-forming elements were available in proportions close to those found in chondrites with the classic exception of Ca, whose abundance in the meteor column is much lower as a consequence of being usually forming part of refractory minerals that experience incomplete vaporization (Trigo-Rodríguez et al. 2003).

### 4.4 Terminal mass and meteorite survival

Two of the events discussed here ("La Carlota" and "Doñana") penetrated deep enough in the atmosphere to take into consideration the likely survival of a fraction of the mass in the parent meteoroid. The meteoroid mass surviving the ablation process, $m_E$, can be obtained from the following relationship (Ceplecha et al. 1983):

$$m_E = \left( \frac{1.2 \rho_E \, v_E^2}{(dv/dt)_E \, \rho_m^{2/3}} \right) \tag{4}$$

with $v_E$ and $(dv/dt)_E$ being, respectively, the velocity and deceleration at the terminal point of the luminous trajectory. $\rho_m$ is the meteoroid density and $\rho_E$ the air density at the terminal height. This equation assumes that the surviving mass is spherical in shape.

In fact, the low altitude of the terminal point (25.6±0.5 km) of the SPMN180910 "La Carlota" bolide suggests that this sporadic fireball was a potential meteorite-dropping event. At the end of its luminous phase, the velocity of the bolide was of about 6.7 km s[-1] and the calculation of the terminal mass of the meteoroid yields 40±30 g. Despite this mass is small, the landing point of the meteorite was estimated with our AMALTHEA software, which models the dark flight of the meteoroid by following the method described in (Ceplecha 1987). Atmospheric data provided by AEMET (the Spanish meteorological agency) were used to take into account wind effects. The particle was assumed to be spherical. According to this analysis, the likely landing point would be located around the geographical coordinates 37.6769 º N, 4.9162 º W. This corresponds to a cereal growing area at about 1.5 km from the village after which the fireball was named. Yet, no meteorites were found.

On the other hand, the luminous phase of the SPMN250112 "Doñana" fireball ended at $26.4\pm0.5$ km above the ground level. According to the calculated atmospheric trajectory, at that point its velocity was around 3.1 km s$^{-1}$. The terminal mass obtained from Eq. (4), however, is of the order of 0.1 grams. So, we concluded that the survival of meteorites was not very likely in this case.

## 4.5. Implications for Jupiter family cometary structure and their delivery of meteorites to Earth.

The inferred physical behaviour, tensile strength and bulk chemistry of the two bolides dynamically linked with JFCs (SPMN180910 and SPMN250112) suggest that some materials released from this cometary family might be similar to (or identical to) carbonaceous chondrites. Our results are also suggesting that this cometary family is a potential meteorite-dropping source as previously noted (di Martino and Cellino 2004, Gounelle et al. 2006). It is thought that the JFC are probably fragments of Kuiper Belt Objects (Morbidelli 2008), and this is also suggested from its size distribution (Snodgrass et al. 2011). The Kuiper Belt contains a reservoir of ice-rich bodies that has preserved dynamical clues about early Solar System processes that have set up its current structure (Morbidelli 2008, Jewitt 2008). The current Nice model of planetary evolution has demonstrated to be capable of reproducing many features of our planetary system (Gomes et al. 2005). In general it is thought that every comet experienced different evolutionary histories, and this particular source of comets could be directly associated with bodies scattered from the outer main belt about 3.9 Gyr ago (Morbidelli et al. 2005, 2010). The inwards migration of Jupiter and Saturn produced a massive scattering of bodies that could be injected in different cometary regions like the Kuiper Belt. Then, we think that this evolutionary model might predict naturally the presence of high-strength chondritic bodies in different populations, like e.g. the JFC region. On the other hand, it is also true that bodies in such populations could have been subjected to significant collisional histories. So their surfaces could be heavily brecciated, and their interiors could contain compressed materials with higher strengths than usually expected for comet-forming materials (Trigo-Rodríguez & Blum 2009). And again the presence of high-strength materials from collisionally evolved comets fits well the recently revisited scenario on the higher collisional rates suffered by cometary families (Bottke et al. 2005, 2007, 2012).

Additionally, it has been recently found that the amount of shock experienced by carbonaceous chondrites has direct effect on the bulk density of these materials (Macke et al. 2011). It is also plausible that these materials trace the remains of disrupted comets as was previously envisioned in a scenario proposed thanks to the observation of a superbolide produced by a 1 m sized meteoroid dynamically associated with the disappeared comet C/1919 Q2 Metcalf (Trigo-Rodríguez et al. 2009b). It is remarkable that, due to the intrinsically high relative velocity of JFCs colliding with other bodies, many of these collisions could be catastrophic, producing fragments like the ones presented in this paper. From an astrobiological point of view it is remarkable that the discovery of ocean-like water in 103P/Hartley 2 (Hartough et al. 2011) opens the possibility that JFC meteoroids could have been an important source of water and organics to Earth. This source is not only associated with sporadic events as those presented here, since there are several important active meteoroid streams associated with JFCs (Williams 2009). In fact, dynamically these streams are driven by the gravitational perturbation of Jupiter. Consequently, we think that there is growing evidence that JFCs can be formed, at least partially, by high-strength materials of chondritic nature. We envision that those reaching the top atmosphere at grazing angles and with moderate velocities (in other words, with the optimal geometrical conditions) could be a source of meteorites, but also of ablation-generated volatiles to Earth's atmosphere (Trigo-Rodríguez 2013).

## 5 CONCLUSIONS

We have described and analyzed four bright fireballs observed over Spain from 2010 to 2012. Our main conclusions are:

1) Except for the SPMN150812 "Torrecera" κ-Cygnid fireball, which ended its luminous phase at an altitude of around 80 km above the ground level, the rest of the events analyzed here were deep-penetrating bolides. In particular, the terminal points of the SPMN180910 "La Carlota" and SPMN 250112 "Doñana" sporadic fireballs were located at a height of about 25.6 and 26.4 km, respectively.

2) The meteoroids that produced the sporadic "Doñana" and "La Carlota" fireballs exhibited a relatively high strength. This is particularly true for the parent meteoroid of the SPMN180910 "La Carlota" event, which exhibited values close to those found for stony meteorites. The values obtained for the κ-Cygnid and the Geminid events are very close to those previously reported in the literature for other members of these meteoroid streams.

3) The calculation of the orbital elements show that, except for the "Peñaflor" Geminid, the parent meteoroid of these events followed Jupiter-Family Comet (JFC) orbits before impacting the Earth. So, our results support the idea of high-strength meteoroids moving in JFC orbits.

4) The analysis of the atmospheric trajectory shows that the SPMN180910 "La Carlota" fireball was a potential meteorite-dropper. However, the calculated mass of the surviving material is below 100 grams.

5) The analysis of the intensity of the emission lines of the three fireballs with available spectra suggests that the bolides "Peñaflor", "Torrecera" and "Doñana" were produced by meteoroids of chondritic nature.

6) From an astrobiological point of view, the continuous delivery of high-strength and large JFC meteoroids could place this family as an important source of water and organics to Earth.

## **ACKNOWLEDGEMENTS**

We acknowledge support from the Spanish Ministry of Science and Innovation (project AYA2011-26522) and Junta de Andalucía (project P09-FQM-4555). We also thank the AstroHita Foundation for its continuous support in the operation of the meteor observing station located at La Hita Astronomical Observatory.

# TABLES

Table 1. Geographical coordinates of the meteor observing stations involved in this work.

| Station # | Station name | Longitude (W) | Latitude (N) | Alt. (m) |
|-----------|--------------|---------------|--------------|----------|
| 1 | Sevilla | 5º 58' 50" | 37º 20' 46" | 28 |
| 2 | La Hita | 3º 11' 00" | 39º 34' 06" | 674 |
| 3 | Huelva | 6º 56' 11" | 37º 15' 10" | 25 |
| 4 | Sierra Nevada | 3º 23' 05" | 37º 03' 51" | 2896 |
| 5 | El Arenosillo | 6º 43' 58" | 37º 06' 16" | 40 |
| 7 | Molina de Segura | 1º 09' 50" | 38º 05' 54" | 94 |

Table 2. Absolute magnitude (M), trajectory and geocentric radiant data (J2000) for the fireballs analyzed in this work. The values of the beginning ($H_b$), ending ($H_e$) and maximum brightness ($H_{max}$) heights are indicated. $V_\infty$, $V_g$ and $V_h$ are the observed preatmospheric, geocentric and heliocentric velocity, respectively.

| SPMN Code and name | Date | Time (UTC) ±0.1s | M | $H_b$ (km) | $H_{max}$ (km) | $H_e$ (km) | $\alpha_g$ (º) | $\delta_g$ (º) | $V_\infty$ (km s$^{-1}$) | $V_g$ (km s$^{-1}$) | $V_h$ (km s$^{-1}$) |
|---|---|---|---|---|---|---|---|---|---|---|---|
| 180910 "La Carlota" | Sep. 18, 2010 | 20h04m27.0s | -11.0 ±0.5 | 85.8 ±0.5 | 51.3 ±0.5 | 25.6 ±0.5 | 298.5 ±0.7 | 46.4 ±0.3 | 21.7 ±0.3 | 18.7 ±0.3 | 38.7 ±0.3 |
| 250112 "Doñana" | Jan. 25, 2012 | 20h20m07.3s | -10.1 ±0.5 | 78.4 ±0.5 | 37.2 ±0.5 | 26.4 ±0.5 | 34.5 ±0.3 | -3.8 ±0.3 | 14.7 ±0.3 | 9.9 ±0.4 | 39.8 ±0.4 |
| 150812 "Torrecera" | Aug. 15, 2012 | 23h44m39.2s | -10.2 ±0.5 | 104.5 ±0.5 | 80.0 ±0.5 | 80.0 ±0.5 | 291.3 ±0.3 | 60.6 ±0.3 | 27.1 ±0.3 | 24.8 ±0.3 | 37.6 ±0.3 |
| 121212 "Peñaflor" | Dec. 12, 2012 | 3h47m19.7s | -10.0 ±0.5 | 101.6 ±0.5 | 65.3 ±0.5 | 39.9 ±0.5 | 114.8 ±0.3 | 34.9 ±0.3 | 39.0 ±0.3 | 37.5 ±0.3 | 34.9 ±0.3 |

Table 3. Orbital elements (J2000) and Tisserand parameter with respect to Jupiter ($T_J$) for the bolides discussed in the text.

| SPMN Code and name | a (AU) | e | i (°) | Ω (°) | ω (°) | q (AU) | $T_J$ |
|---|---|---|---|---|---|---|---|
| 180910 "La Carlota" | 3.4±0.2 | 0.71±0.02 | 26.4±0.3 | 175.6521±10⁻⁴ | 205.0±0.6 | 0.965±0.001 | 2.5±0.1 |
| 250112 "Doñana" | 4.1±0.6 | 0.76±0.04 | 4.0±0.1 | 125.0826±10⁻⁴ | 357.0±0.3 | 0.9839±0.0001 | 2.4±0.1 |
| 150812 "Torrecera" | 2.6±0.1 | 0.62±0.01 | 40.8±0.4 | 143.3539±10⁻⁴ | 199.8±0.3 | 0.9895±0.0007 | 2.8±0.1 |
| 121212 "Peñaflor" | 1.52±0.04 | 0.920±0.003 | 36.2±0.8 | 260.3488±10⁻⁴ | 326.0±0.5 | 0.120±0.003 | 3.7±0.1 |

Table 4. Photometric mass ($m_p$) and estimated diameter (D) of each meteoroid.

| SPMN Code and name | $m_p$ (kg) | D (cm) (d=2.4 g cm⁻³) | D (cm) (d=3.7 g cm⁻³) |
|---|---|---|---|
| 180910 "La Carlota" | 1.9±0.2 | 11.4±0.3 | 9.9±0.3 |
| 250112 "Doñana" | 5.7±0.6 | 16.5±0.6 | 14.3±0.5 |
| 150812 "Torrecera" | 0.18±0.02 | 5.2±0.2 | 4.5±0.2 |
| 121212 "Peñaflor" | 0.12±0.01 | 4.5±0.1 | 3.9±0.1 |

Table 5. Aerodynamic pressure for flares and break-up processes discussed in the text.

| SPMN Code and name | Flare # | Height (km) | Velocity (km s⁻¹) | Aerodynamic pressure (dyn cm⁻²) |
|---|---|---|---|---|
| 180910 | 1 | 51.3±0.5 | 19.6±0.3 | (3.2±0.4)·10⁶ |
| "La Carlota" | 2 | 38.1±0.5 | 16.9±0.3 | (1.4±0.4)·10⁷ |
| 250112 | 1 | 43.1±0.5 | 12.1±0.3 | (3.6±0.4)·10⁶ |
| "Doñana" | 2 | 37.2±0.5 | 9.8±0.3 | (5.6±0.4)·10⁶ |
| 150812 "Torrecera" | 1 | 80.0±0.5 | 26.1±0.3 | (1.1±0.4)·10⁵ |
| 121212 | 1 | 65.3±0.5 | 37.4±0.3 | (2.0±0.4)·10⁵ |
| "Peñaflor" | 2 | 57.1±0.5 | 35.1±0.3 | (5.1±0.4)·10⁵ |

Table 6. Relative abundances with respect to Si of the main rock-forming elements derived from the analysis of emission spectra.

| SPMN Code and name | T (K) ±100 | N (cm$^{-2}$) | Mg | Na | Fe | Ca | Mn (×10$^{-4}$) |
|---|---|---|---|---|---|---|---|
| 250112 "Doñana" | 4,200 | $1\cdot10^{14}$ | 0,77 | 0,05 | - | 0,023 | - |
| 150812 "Torrecera" | 4,500 | $1\cdot10^{15}$ | 0.80 | 0,05 | - | 0,028 | 25 |
| 121212 "Peñaflor" | 4,800 | $5\cdot10^{15}$ | 1.03 | 0,04 | 0,82 | 0,026 | - |
| 1P/Halley | - | - | 0.54 | 0.054 | 0.280 | 0.034 | 30 |
| IDPs | - | - | 0.85 | 0.085 | 0.63 | 0.048 | 150 |
| CI chondrites | - | - | 1.06 | 0.060 | 0.90 | 0.071 | 90 |
| CM chondrites | - | - | 1.04 | 0.035 | 0.84 | 0.072 | 60 |

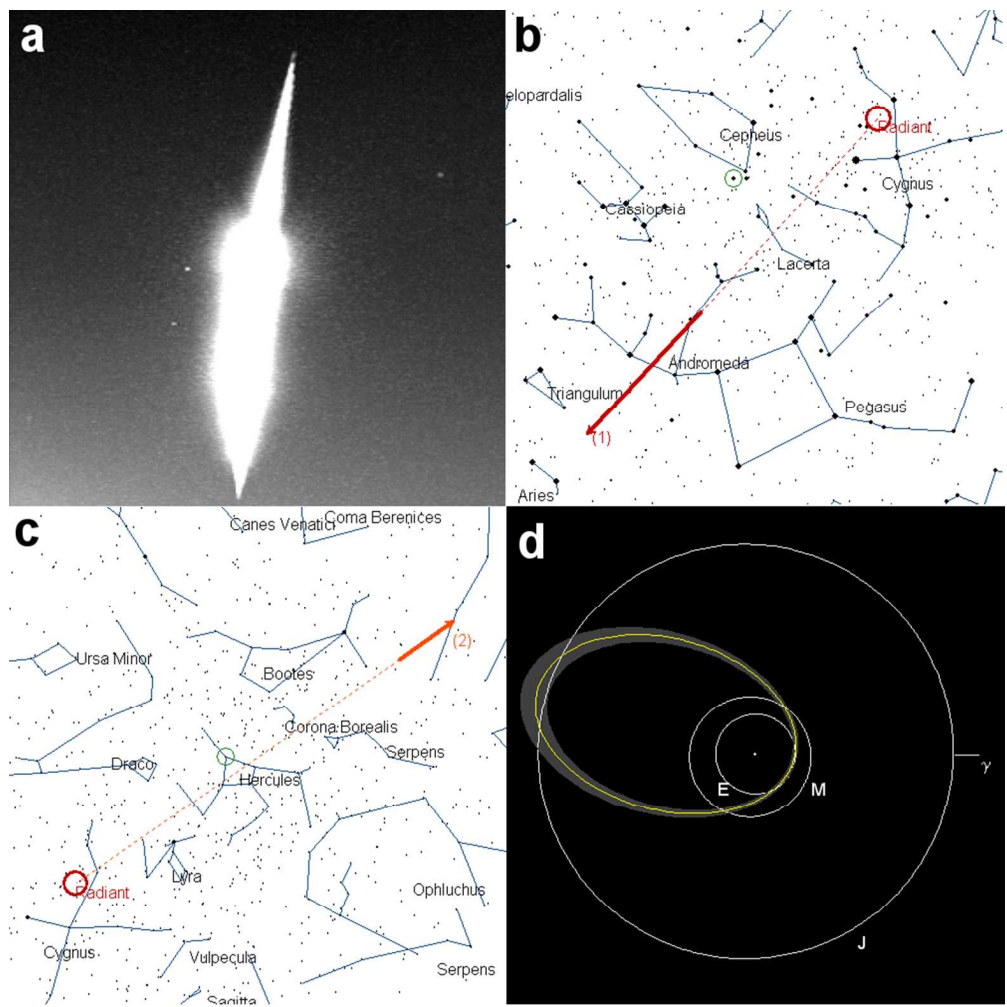

Figure 1. a) Composite image of the sporadic "La Carlota" fireball (code SPMN180910), imaged on Sep. 18, 2010 at 20h04m27.0±0.1s UTC from Sevilla. b) Apparent trajectory of the bolide as seen from Sevilla and c) La Murta meteor observing stations. d) Heliocentric orbit of the meteoroid projected on the ecliptic plane. The gray area represents the uncertainty in the orbit by taking into account the uncertainty in the semimajor axis.
352x352mm (72 x 72 DPI)

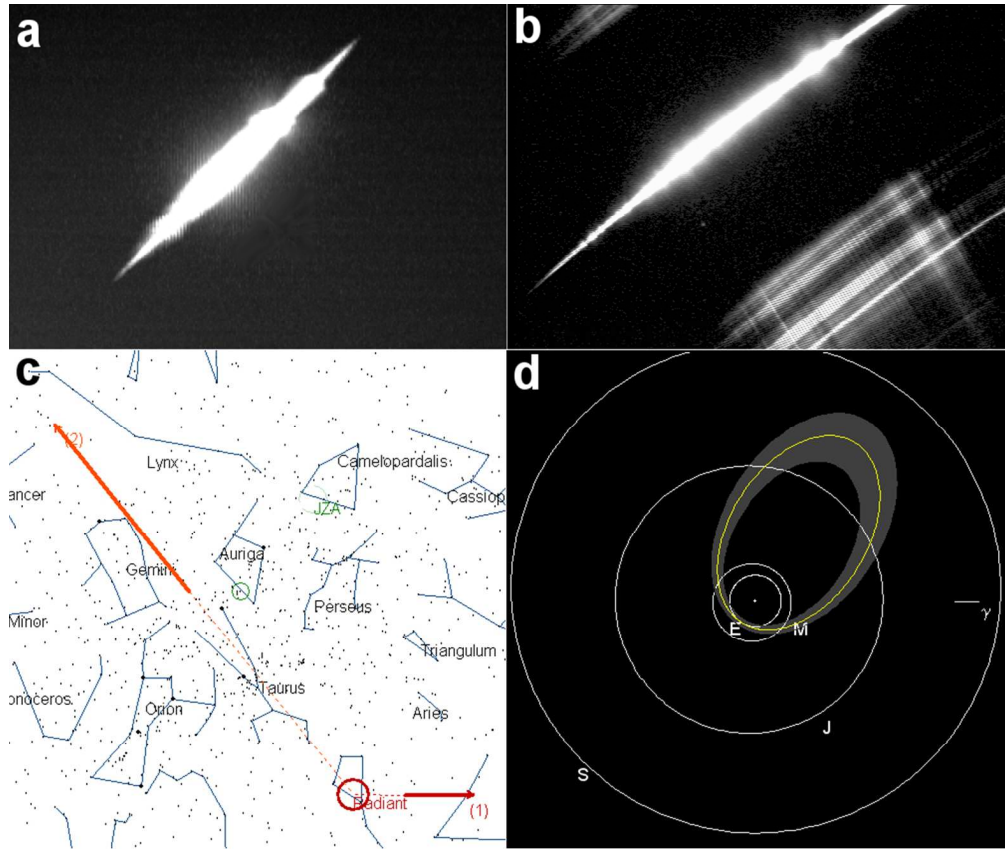

Figure 2. a) Composite image of the "Doñana" fireball (code SPMN250112), imaged on Jan. 25, 2012 at 20h20m07.3±0.1s UTC from a) Sevilla and b) El Arenosillo. c) Apparent trajectory of the bolide as seen from Sevilla (1) and El Arenosillo (2). d) Heliocentric orbit of the meteoroid projected on the ecliptic plane. The gray area represents the uncertainty in the orbit by taking into account the uncertainty in the semimajor axis.

352x299mm (72 x 72 DPI)

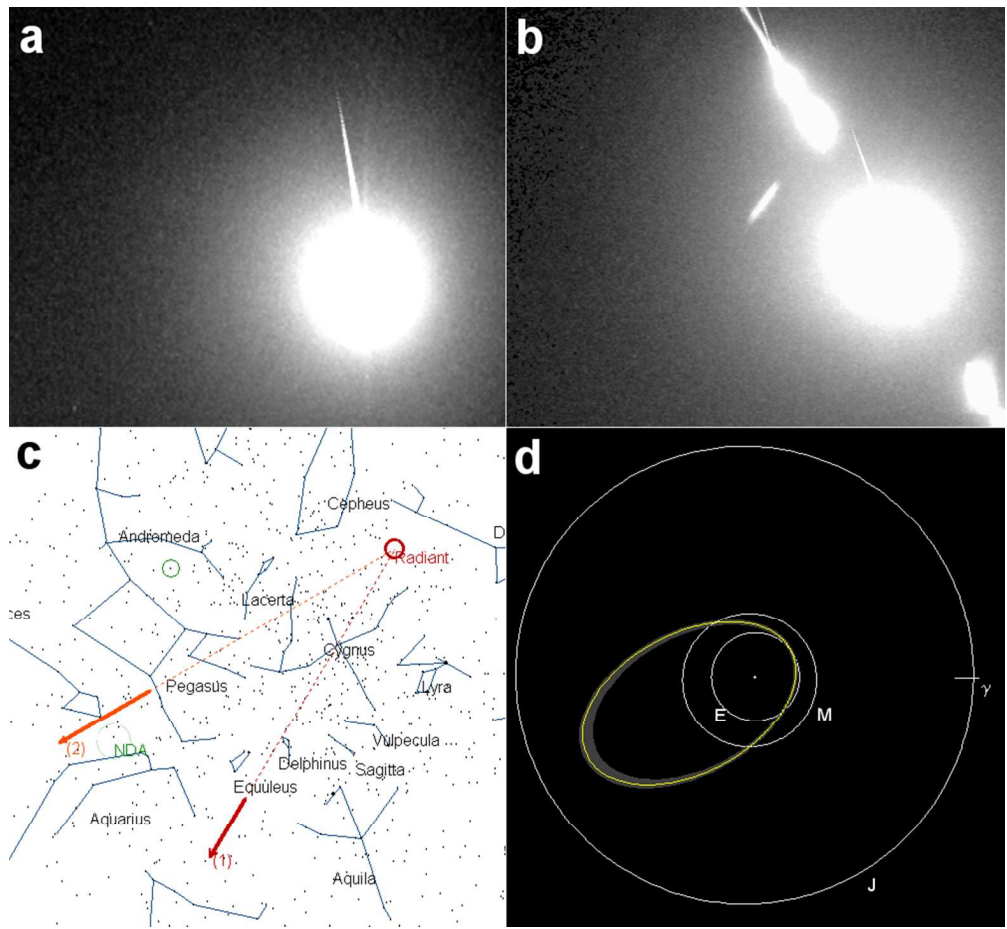

Figure 3. Composite image of the "Torrecera" κ-Cygnid fireball (code SPMN150812), imaged on Aug. 15, 2012 at 23h44m39.2±0.1s UTC from a) Sevilla and b) El Arenosillo. c) Apparent trajectory of the bolide as seen from Sevilla (1) and El Arenosillo (2). d) Heliocentric orbit of the meteoroid projected on the ecliptic plane. The gray area represents the uncertainty in the orbit by taking into account the uncertainty in the semimajor axis.
352x325mm (72 x 72 DPI)

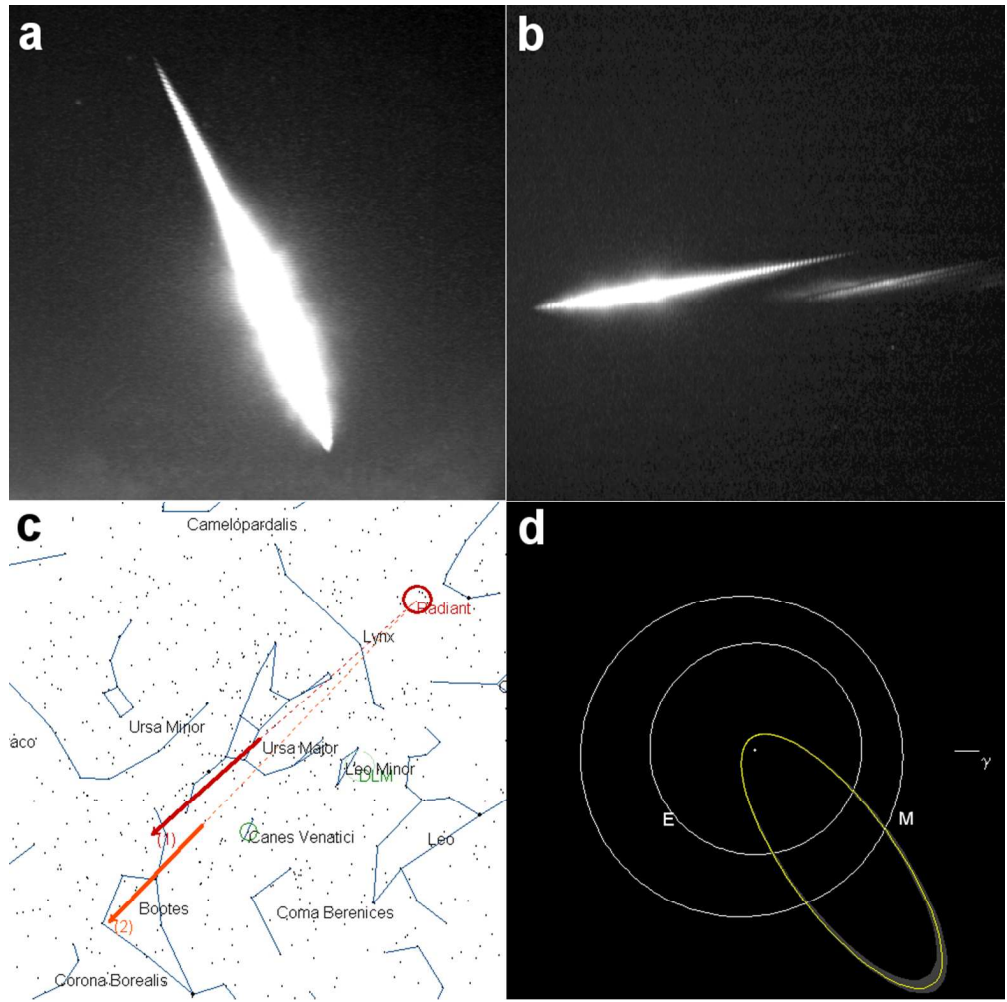

Figure 4. Composite image of the "Peñaflor" Geminid fireball (code SPMN121212), imaged on Dec. 12, 2012 at 3h47m19.7±0.1s UTC from a) Sevilla and b) El Arenosillo. c) Apparent trajectory of the bolide as seen from (1) Sevilla and (2) El Arenosillo. d) Heliocentric orbit of the meteoroid projected on the ecliptic plane. The gray area represents the uncertainty in the orbit by taking into account the uncertainty in the semimajor axis.
352x352mm (72 x 72 DPI)

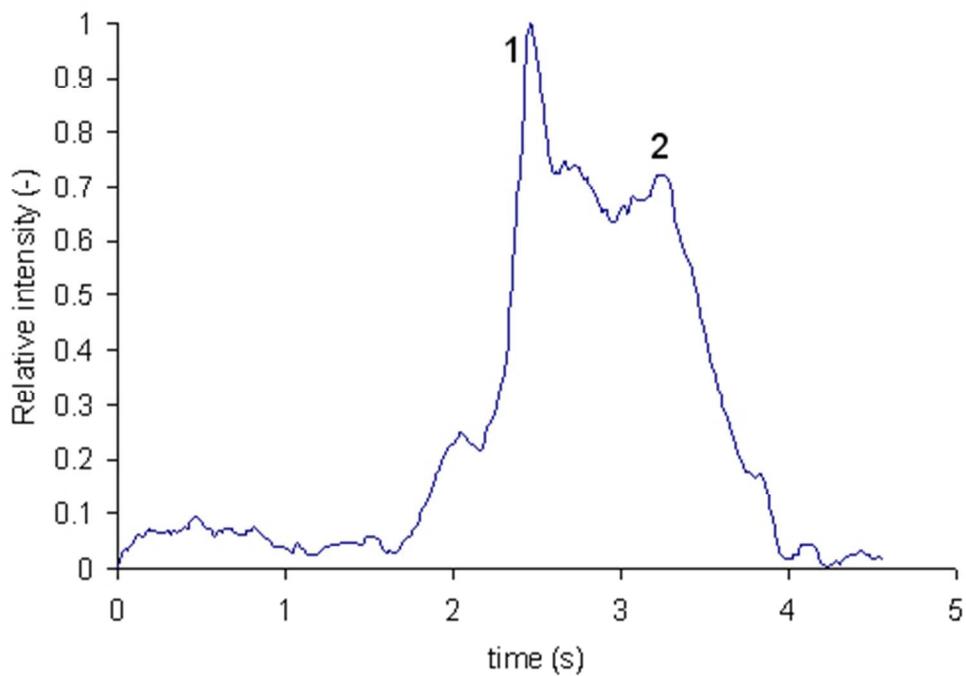

Figure 5. Light curve (relative pixel intensity vs. time) of the SPMN180910 "La Carlota" fireball.
178x127mm (72 x 72 DPI)

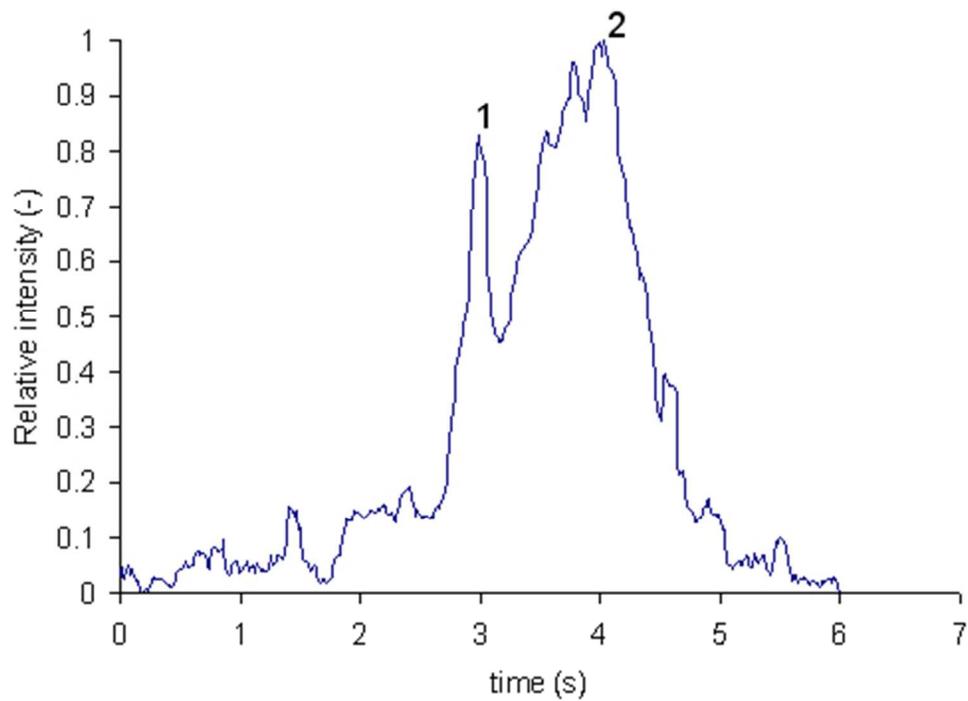

Figure 6. Light curve (relative pixel intensity vs. time) of the SPMN250112 "Doñana" fireball.
177x128mm (72 x 72 DPI)

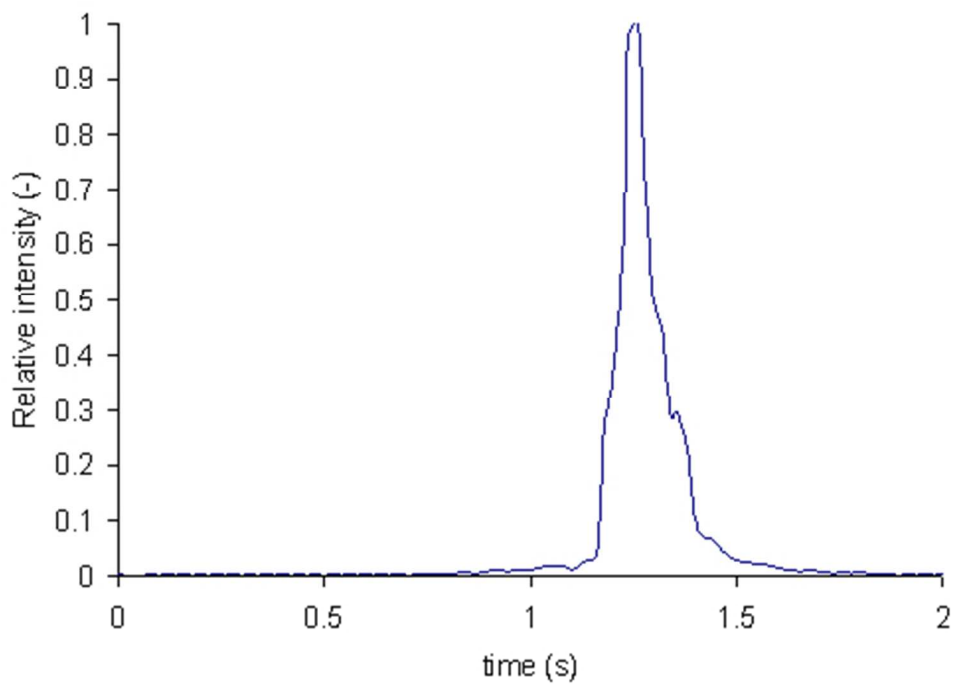

Figure 7. Light curve (relative pixel intensity vs. time) of the SPMN150812 "Torrecera" κ-Cygnid bolide.
177x128mm (72 x 72 DPI)

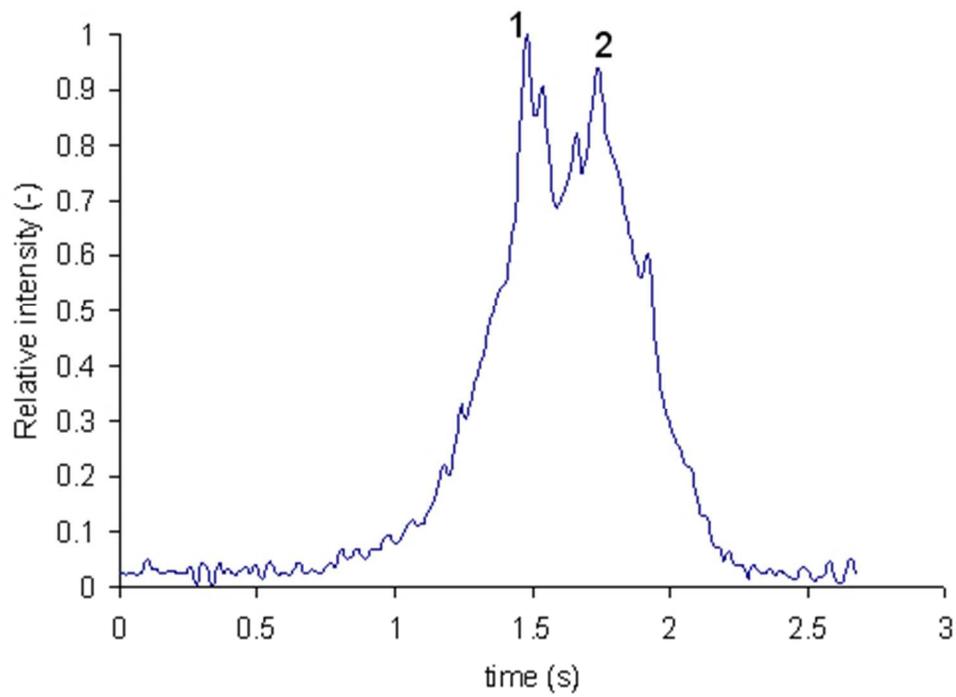

Figure 8. Light curve (relative brightness vs. time) of the SPMN121212 "Peñaflor" fireball. Main fulgurations for which the aerodynamic pressure was calculated are highlighted.
177x128mm (72 x 72 DPI)

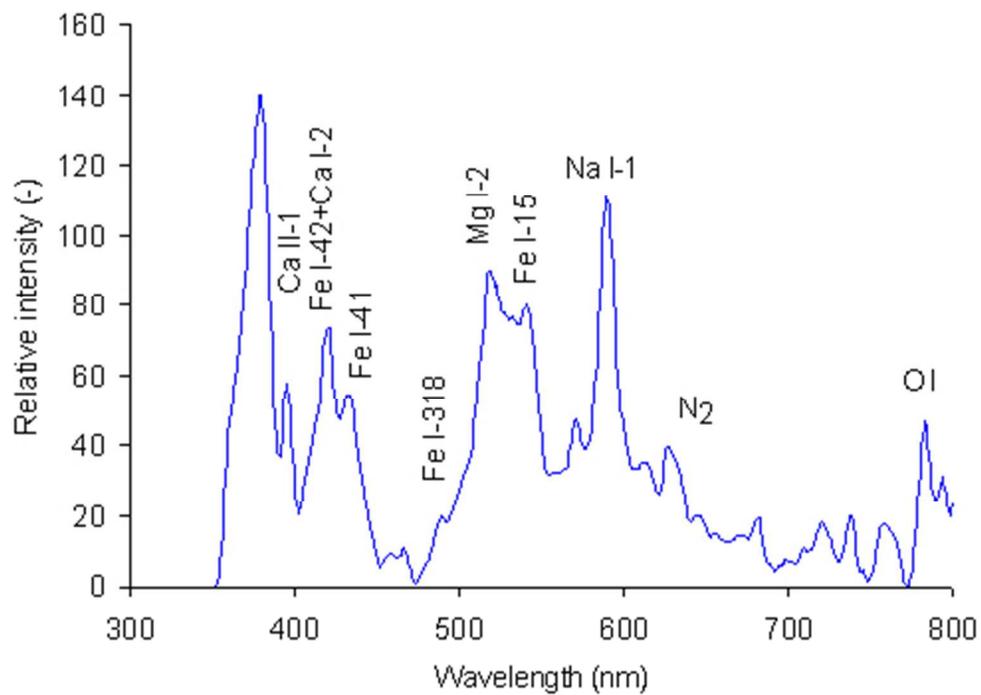

Figure 9. Calibrated emission spectrum recorded for the SPMN250112 "Doñana" fireball. Intensity is expressed in arbitrary units.
177x127mm (72 x 72 DPI)

1
2
3
4
5
6
7
8
9
10
11
12
13
14
15
16
17
18
19
20
21
22
23
24
25
26
27
28
29
30
31
32
33
34
35
36
37
38
39
40
41
42
43
44
45
46
47
48
49
50
51
52
53
54
55
56
57
58
59
60

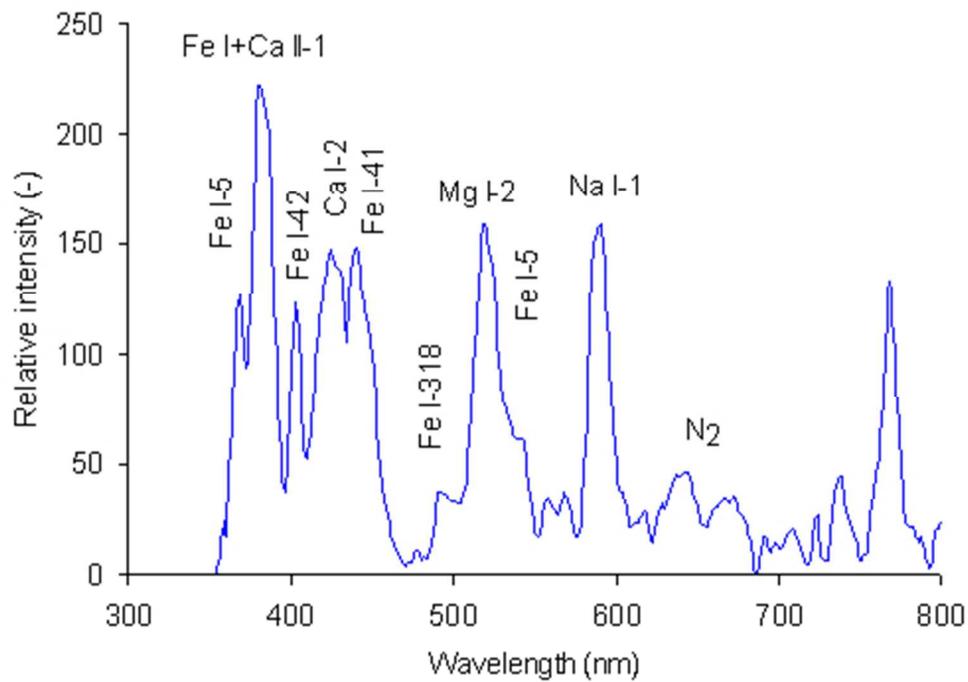

Figure 10. Calibrated emission spectrum recorded for the SPMN150812 "Torrecera" κ-Cygnid bolide.
Intensity is expressed in arbitrary units.
178x128mm (72 x 72 DPI)

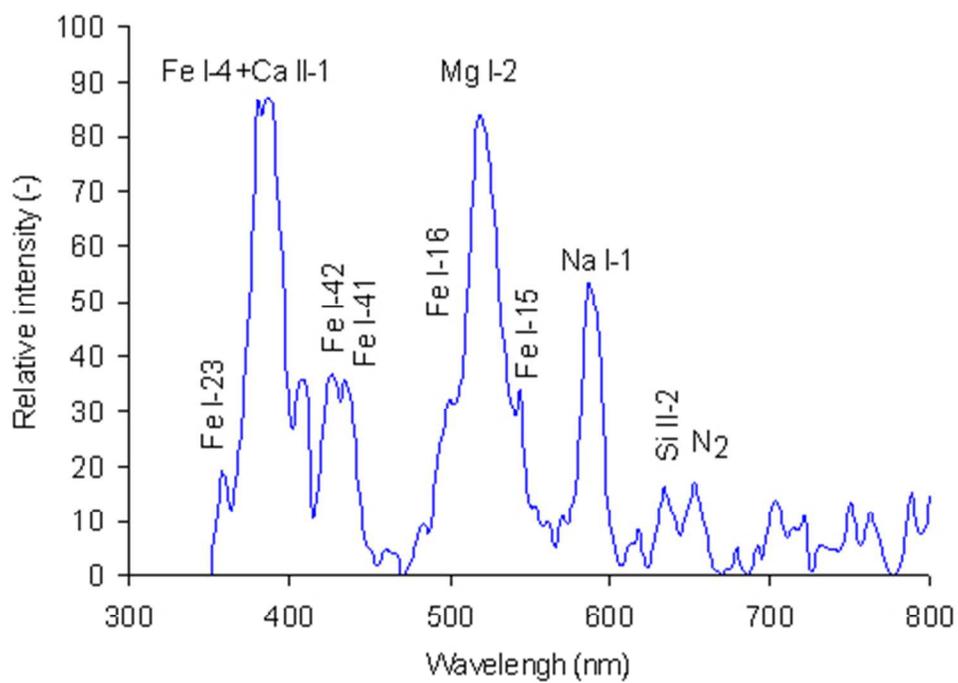

Figure 11. Calibrated emission spectrum recorded for the SPMN121212 "Peñaflor" Geminid bolide. Intensity is expressed in arbitrary units.
177x128mm (72 x 72 DPI)